\documentclass{article}

    \PassOptionsToPackage{numbers, compress}{natbib}

\usepackage[final]{neurips_2020}

\usepackage[utf8]{inputenc} 
\usepackage[T1]{fontenc}    
\usepackage{hyperref}       
\usepackage{url}            
\usepackage{booktabs}       
\usepackage{amsfonts}       
\usepackage{nicefrac}       
\usepackage{microtype}      
\usepackage{graphicx}       
\usepackage{xcolor}

\graphicspath{ {./Figures/} }

\title{Ontological Learning from Weak Labels}

\author{%
  Larry Tang \\
    ECE Department\\
    Carnegie Mellon University \\
    \texttt{lawrenct@andrew.cmu.edu} \\
    \And
    Po Hao Chou \\
    ECE Department\\
    Carnegie Mellon University \\
    \texttt{pohaoc@andrew.cmu.edu} \\
    \And
    Yi Yu Zheng \\
    ECE Department\\
    Carnegie Mellon University \\
    \texttt{yiyuz@andrew.cmu.edu} \\
    \And
    Ziqian Ge\\
    ECE Department\\
    Carnegie Mellon University \\
    \texttt{ziqiang@andrew.cmu.edu} \\
    \And 
    Ankit Shah \\
    LTI Department\\
    Carnegie Mellon University\\
    \texttt{aps1@andrew.cmu.edu} \\
    \And
    Bhiksha Raj\\
    LTI Department\\
    Carnegie Mellon University\\
    \texttt{bhiksha@cs.cmu.edu} \\
}

\begin{document}

\maketitle

\begin{abstract}
  Ontologies encompass a formal representation of knowledge through the definition of concepts or properties of a domain, and the relationships between those concepts. 
  In this work, we seek to investigate whether using this ontological information will improve learning from weakly labeled data, which are easier to collect since it requires only the presence or absence of an event to be known.  We use the AudioSet ontology and dataset, which contains audio clips weakly labeled with the ontology concepts and the ontology providing the "Is A" relations between the concepts.  We first re-implemented the model proposed by \cite{soundevent_ontology} with modification to fit the multi-label scenario and then expand on that idea by using a Graph Convolutional Network (GCN) to model the ontology information to learn the concepts. We find that the baseline Siamese does not perform better by incorporating ontology information in the weak and multi-label scenario, but that the GCN does capture the ontology knowledge better for weak, multi-labeled data. In our experiments, we also investigate how different modules can tolerate noises introduced from weak labels and better incorporate ontology information. Our best Siamese-GCN model achieves mAP=0.45 and AUC=0.87 for lower level concepts and mAP=0.72 and AUC=0.86 for higher level concepts, which is an improvement over the baseline Siamese but about the same as our models that do not use ontology information.

\end{abstract}

\section{Introduction}
Ontologies represent hierarchical concepts in our brains through categories and relationships of domain knowledge. For example when we hear a sound, even if we don't recognize the specific animal species making the sound, we still recognize that given sound is a type of animal sound. Like humans, a machine can utilize the ontology information to help to classify an object it hasn't seen before as a higher level label. Incorporating these hierarchical relations from ontologies can improve classification of semantically different observations that appear similar or provide more general descriptors of ambiguous subclasses.  Many recent works have looked into how to use this external knowledge to extract better feature representations as well as how to embed the knowledge into model architectures.

In this paper, we investigate whether the learning of ontological classes from weak labeled data can be improved by including external knowledge of these relationships. For example, an audio clip may contain a \textit{dog howling}, a \textit{baby crying}, and an \textit{engine idling}, but is weakly labeled simply as a \textit{dog howling} is present. We will predict the ontological information of that data point, which includes \textit{animal sounds} and \textit{living things} as the higher level concepts.  The ontology knowledge base can be also be provided or incorporated into the neural network to aid in the classification task.  One of the challenges in this paper is that the network should also be able to notice that there may be parts of the data that could belong to other ontological labels, or could be tagged in the same higher level ontological label but may not be labeled as such.  For example, a piece of an audio clip containing both \textit{dog howling} and \textit{cat meowing} may only be weakly labeled as \textit{dog howling}, while both \textit{dog howling} and \textit{cat meowing} could lead to the prediction of the higher level ontological label \textit{animal sounds} or \textit{living thing}. We would like to look into whether the network can use that hidden knowledge to improve the prediction for respective ontological categories. 

The advantage of weakly labeled datasets is that weak labels are easier to collect since they require only the presence or absence of an event to be known, which leads to better scalability as demonstrated by Google's AudioSet \cite{audiosetdata}.  However, they may introduce noise into the labels which makes it difficult for models to learn from.  Investigating this problem is thus important because it may lead to improved performance for classification tasks in which it is difficult to obtain strongly labeled datasets.  We look to improve upon some previous work in the space of ontology prediction, hierarchical learning, and knowledge graphs by implementing two different models which embed the ontology information to predict the ontology classes and hierarchies from weak labels. We hope to gain a better understanding of ontological embeddings, whether they can improve classification in the weak label scenario, and the advantages or disadvantages of using weak labels. The following sections give a more detailed background of recent literature and then discusses the techniques and models we implement and our final results and conclusions.


\section{Related Work}

Domain knowledge can be represented in different ways but here we focus on ontology based or knowledge graph representations which can capture broader relations between concepts that are not only hierarchical. There are many recent works investigating how to embed domain knowledge to improve classification as well as the prediction of the actual ontology classes themselves.   One line of research has been on how to use these hierarchical concepts to improve feature embeddings, while another has focused on how to incorporate this knowledge into the actual network architecture. The following paragraphs discuss recent surveys and work on these two methods.

\paragraph{Feature Extraction using Ontologies}
In a recent survey paper \cite{featureselection}, they discuss how the use of ontologies for feature selection have focused on applications in text classification. One recent work uses the Wordnet taxonomy \cite{WordNet} as an ontology to compute a similarity measurement that determines which features to keep or discard.  Other works have used ontologies to map text documents to a more general concept hierarchy, essentially representing the content of the text. Some other fields have even used building types as concepts and medical ontologies to identify features for drug classification.

The embeddings trained from ontologies are also used to improve some specific learning tasks such as Zero Shot Learning. In \cite{onto_zeroshot}, the author defined an ontology schema as an interface for knowledge graphs such as WordNet \cite{WordNet} and NELL \cite{NELL}. They use the ontology schema as an input with some noise to a Generative Adversarial Networks to improve Zero Shot Learning.

Another recent survey paper related to the use of ontology, \cite{The_use_of_ontologies}, discusses how ontologies allow an interaction between data stored in different formats and can further be used as the basis to guide and validate models of some specific domains. Based on this prospect, future research can be focused on an investigation of the feasibility to use ontological knowledge base as a foundation to efficiently discover useful information to assist analysis. Such research efforts include: (1) To what extent the data can be extracted and aggregated and (2) How can the extracted data as well as domain knowledge be represented.  All of these techniques focus on how to use the ontology knowledge to extract better feature representations that model the ontology information.

\paragraph{Classification using Ontologies}
There have been some works trying to apply ontology to mitigate the label noises in audio data. In \cite{MT-GCN}, a Multi-task Learning based Graph Convolutional Network (MT-GCN) is proposed to utilize ontology information to regularize multi-label audio tagging problem. The model is an extension from the network proposed by \cite{ML-GCN}, which uses GCN \cite{GCN} to learn the dependency between multiple labels in images. Instead of using the label dependency as correlation graph, MT-GCN chooses to use the ontology hierarchy information to construct the correlation graph. To be more specific, it represents node by ontology class and builds their correlation according to the parent child relation or connectionism. It then uses the Binary Cross Entropy Loss information from Audio Tagging tasks to train the embedding of ontology class.

On the subject of prediction of ontological labels, \cite{retrievingframework} by G. Wichern et al constructed an ontological framework where sounds are connected to each other based on the similarity between acoustic features, while semantic tags and sounds are connected through link weights that are optimized based on user-provided tags. More specifically, they proposed link weights between different sound samples to be determined using a likelihood-based technique, and estimated a hidden Markov model (HMM) from each of the trajectories of features. They have also proposed link weights between concepts using vector metrics, a similarity metric from the WordNet::Similarity library, and link weights between concepts and sounds by computing KL divergence between the output of the ontological framework and a votes matrix of tags, and from that constructed an integrated system for text-based retrieval of unlabeled audio, content-based query-by-example, and automatic annotation of unlabeled sound files. Their approach of relating sounds with similar features together is heuristic for learning higher level ontological labels.


\section{Methods}

\subsection{Dataset and Ontology}

\begin{figure}[ht]
    \centering
    \includegraphics[scale=0.5]{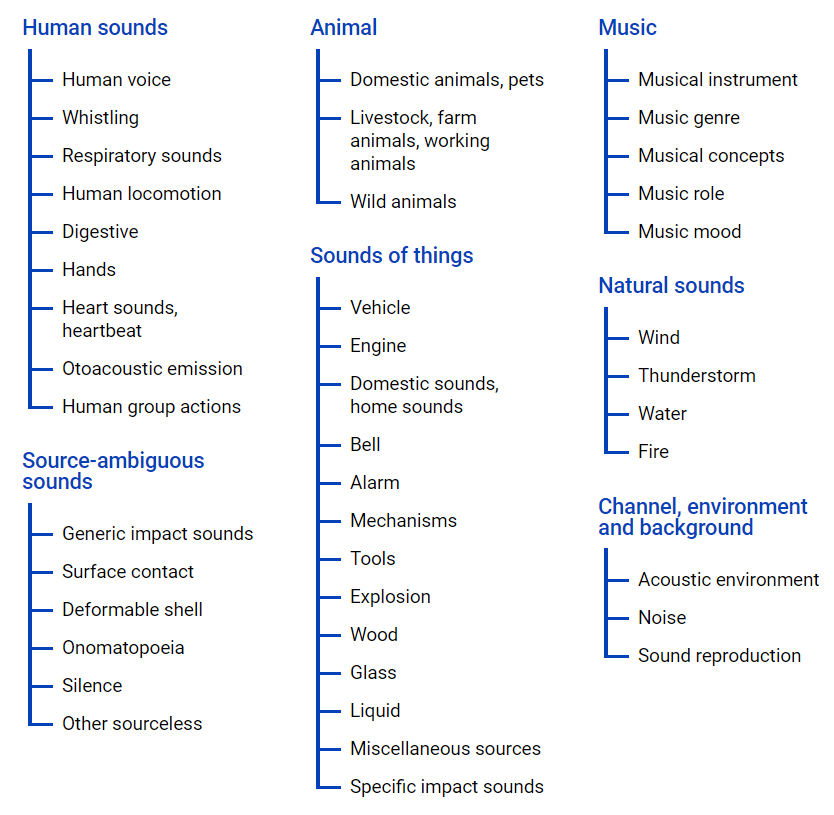}
    \caption{The top two highest level classes in the AudioSet ontology}
    \label{fig:audioset_ontology}
\end{figure}

In this paper, we use the AudioSet dataset, which consists of an ontology of 632 audio event classes and 2,084,320 human-labeled 10-second sound clips drawn from YouTube videos.  The 128-dimensional Audioset features and ontology files are made available by Google and can be easily accessed online from previous work \cite{shah2018closer}, \cite{audiosetdata}.  The dataset is a massive compilation of 10 second clips from various YouTube videos which have been weakly labeled. The feature embeddings provided by \cite{audiosetdata} are extracted from a VGGish model followed by a PCA to reduce down to 128 dimensions and each ten second clip is represented by ten 128-dimension feature vectors, i.e. one feature vector represents one second of one clip.  The entire ten second clip can contain multiple labels out of the total 527 classes in the Audioset data.  The classes encompass a wide variety of sounds such as human noises, music, animals, or vehicles. The first two higher level classes are shown in Figure \ref{fig:audioset_ontology}.

We use the training set of Audioset for training our models, which contains about 22,160 ten second clips or about 221,160 feature embeddings.  We then preprocess the corresponding labels of each audio clip to get the labels in the top two levels of the Audioset ontology. The validation set is about 20\% of the training set size and is drawn from the unbalanced training set of Audioset, and the Audioset test set contains about 20,383 audio clips. The same label processing is done for the validation and test set to obtain two levels of ontology labels. Although there are more levels to the Audioset ontology, we choose to focus on the top two levels to fit our model architectures.

\begin{figure}[h]
    \centering
    \includegraphics[scale=0.5]{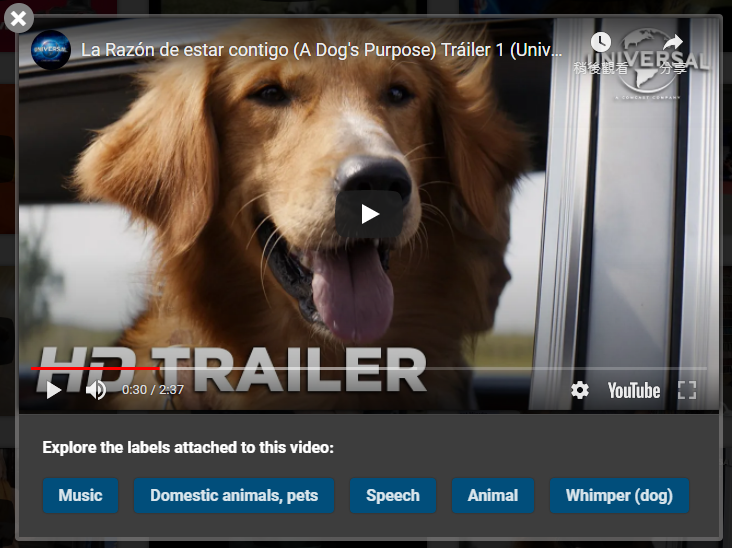}
    \caption{The weak labels in a AudioSet clip \cite{audiosetdata} could be related due to different factors.}
    \label{fig:cute_dog}
\end{figure}

To give a more general idea of the Audioset data, consider a weakly labeled audio clip which could have multiple labels with no timing information, but the labels could be related due to their ontology. For example, a human voice and hands could possibly co-occur because they are both natural sounds. The labels could also be related due to their dependency, such as a human crying could co-occur with sad music. On the other hand the labels could also not be related with each other. For instance, a dog barking sound and car engine sound could co-occur in a clip accidentally. We can observe this from one of the AudioSet clips, such as Figure \ref{fig:cute_dog}.

\subsection{Framework}
We consider multi-labeled training data \{($\mathbf{x}_1$, $\mathbf{y}_1$), ..., ($\mathbf{x}_n$, $\mathbf{y}_n$)\} where $\mathbf{x}_i \in \mathcal{X}$ is a single 128-dimension audio feature representation and the corresponding $\mathbf{y}_i$ is a collection of labels \{($y_1^1$, $y_1^2$, ..., $y_1^{T_1}$), ..., ($y_{k}^1$, $y_{k}^2$, ..., $y_k^{T_k}$)\} with $k$ equal to the number of ontology levels and $T_i$ equal to the number of labels for $\mathbf{x}$ at the $i^{th}$ ontology level. Thus in our post-processed dataset we have $k$ = 2 and will refer to these as subclass or "level 1" labels and superclass or "level 2" labels.

\subsection{MLP Network without Ontology Information} 
To investigate whether incorporating the ontology information is beneficial to use with weakly labeled data, we first need to train a model which uses no ontology information. We implement a simple MLP network with three hidden layers and a final output layer that collectively predicts ontology classes.  More concretely, the final layer has size $\sum_{i=1}^k{T_i}$ to predict all ontology labels.  Each hidden layer is of size 512 followed by a BatchNorm layer, ReLU activation, and a Dropout layer.  This model makes no use of the ontology information to aid in the prediction of the ontology labels, making it a good baseline to compare to our models which will incorporate the ontology information.  The final outputs pass through a final sigmoid activation for the multi-label scenario and the model aims to minimize the binary cross entropy loss.


\subsection{Siamese Model with Ontology-based embeddings and Ontological layer}

\begin{figure}[ht]
    \centering
    \includegraphics[scale=0.4]{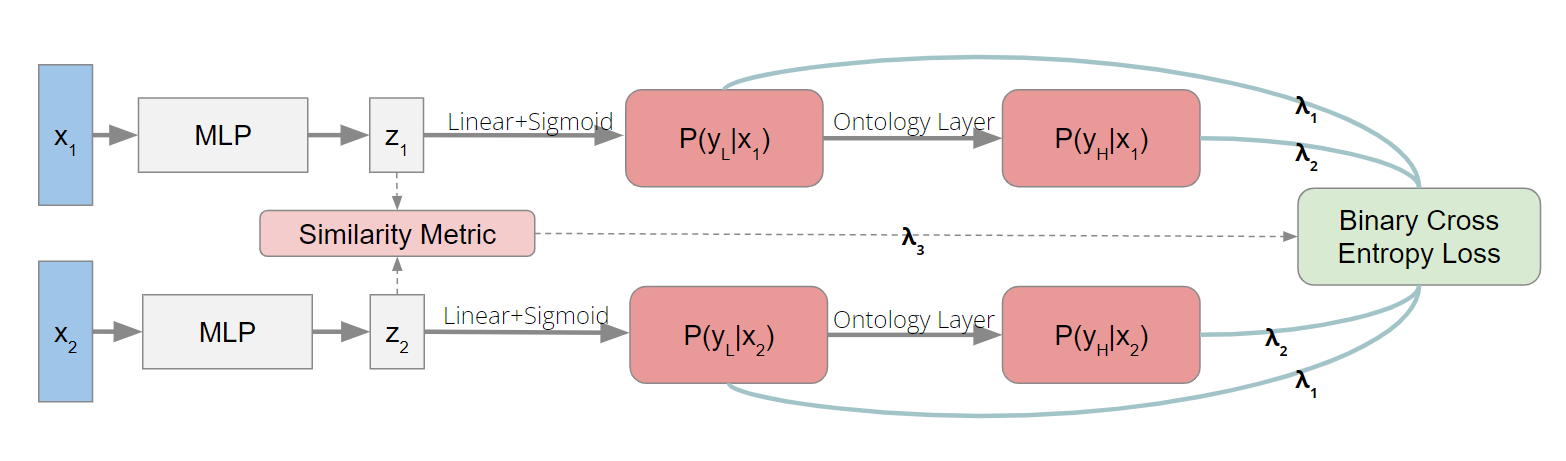}
    \caption{Architecture of Siamese Network + Ontological Layer with modification to fit the multi-label task }
    \label{fig:siamese_ontological}
\end{figure}

The ontology based model from \cite{soundevent_ontology} is a recent contribution that specifically focuses on audio data while also providing a method for prediction of classes at different ontology levels.  Much of the prior work on ontology prediction has been in different areas such as medicine (drug classification) or text classification, whereas \cite{soundevent_ontology} presents a framework that is more relevant and recent. This model will therefore serve as a baseline for the models which will incorporate ontology information. The framework with modification for multi-label scenarios is shown in Figure \ref{fig:siamese_ontological}.

We now present the components of the model and its mathematical formulation, which is an extension of the work in \cite{soundevent_ontology} to learn ontology-based embeddings for classification of multi-labeled data. We use a Siamese neural network (SNN) to separate the ontology-based embeddings by imposing the embedding distance close to 0 if the input pairs are from the same subclass, close to 1 if they are from different sub classes, but the same super class, and close to 2 if they are from different super classes. To fit in the multi-label scenario, we applied two kinds of sampling methods, the first one is that samples fall into the same sub (super) class category if and only if they have exactly the same subclasses; samples fall into different sub (super) class category if any of their sub (super) classes is different. The second approach is that samples fall into the same sub (super) class category if and only if there is any sub (super) class in their intersection; samples fall into different sub (super) class category if there is no any sub (super) class in their intersection.

The base model architecture of the Siamese net is the same MLP with three hidden layers from section 3.3, and each branch of the Siamese net shares the MLP parameters. Given a pair of input vectors $\mathbf{x_1}, \mathbf{x_2},$ the output of the MLP is an ontology embedding $\mathbf{z_1} = f(\mathbf{x_1}), \mathbf{z_2} = f(\mathbf{x_2})$.  The embedding vectors $\mathbf{z_1}, \mathbf{z_2}$ produce subclass probabilities after a sigmoid activation, which is used for the prediction of multi-labeled data.  

The ontological layer, \textbf{M}, then relates the class probabilities of level 1 to those in level 2 of the ontology through the following relation:
\[\mathbf{p}(y_2 | \mathbf{x}) = \mathbf{M} \cdot \mathbf{p}(y_1 | \mathbf{x})\]
We modify the ontological layer proposed in \cite{soundevent_ontology} in order to fit the multi-label data scenario, where \textbf{M} is constructed such that it averages the probabilities of the subclasses within a single superclass. Note that this layer is fixed and not trainable, it depends strictly on the ontology of the training data. 

The final loss function incorporates the binary cross entropy loss of the level 1 classes, $\mathcal{L}_1$, and the level 2 classes, $\mathcal{L}_2$, with the embedding loss, $D_w = (||\mathbf{z_1} - \mathbf{z_2}||_2 - d)^2$ where $d \in \{0, 1, 2\}$ according to the type of input pair.
\[ \mathcal{L} = \lambda_{1} (\mathcal{L}_1^1 + \mathcal{L}_1^2 ) + \lambda_2 (\mathcal{L}_2^1 + \mathcal{L}_2^2 ) + \lambda_3 D_w \]
The base network takes an input feature of dimension 128 before the output layer to the 42 classes in the first ontology level and then through the ontology layer to the 7 classes in the second level.  The baseline paper does not actually report evaluation metrics for the Audioset data, as they focused on other audio datasets: Urban Sounds - US8K and data from the Making Sense of Sounds Challenge. These are both single labeled datasets.

\subsection{Graph Convolutional Network} \label{gcn}

To model both the co-occurrence information introduced from weak labels and the domain knowledge from ontology, we seek to utilize graph embedding approaches to extend the Siamese network architecture. In \cite{MT-GCN} and \cite{ML-GCN}, a Graph Convolution Network (GCN) is shown to be effective for learning useful node representations through the information in a correlation graph. The essential idea is to update the node representations by aggregating information from neighboring nodes.

Following the ideas of previous work, \cite{MT-GCN}, \cite{ML-GCN}, we define the subclass labels and superclass labels as the nodes of the graph and aim to learn the label representation. The knowledge in the graph is encoded as a correlation matrix, which is a crucial part of the GCN. We will describe how it is constructed in Section \ref{gcn_correlation}.

We use one-hot encoding of labels as the initial node representation and use 2 GCN layers to extract embeddings with neighboring information. For each GCN layer we use 2 linear layers to transform the input embedding of the node itself and a graph convolution to aggregate the embeddings from its neighbor.  Given a label embedding $Z \in \mathbb{R^d}^{C \times d}$ (where $C$ is the number of nodes and $d$ is the dimensionality of node features), the graph convolution operations is: 
\[ Z^{l+1} = h(A'Z^lW_1^lW_2^l)\] 
where $W_1^l \in R ^ {d \times d'}$ and $W2^l \in R ^ {d' \times d''}$ are 2 transformation matrices to be learned and $A' \in R^{C \times C}$ is the correlation matrix and $h(.)$ is a non-linear operation which is a LeakyReLU in our experiments.  The dimensions of each the linear layers are 280 and 512 for the first GCN layer and 320 and 128 for the second GCN layer.

\subsubsection{Correlation Matrix} \label{gcn_correlation}

The GCN learns node representations by collecting information from other nodes based on the correlation matrix provided. Thus, how we build the correlation matrix is crucial but also challenging for GCN. In this work, we referred to previous work \cite{MT-GCN, ML-GCN} and experimented on 3 different correlation matrices.

\textbf{Labels Co-occurrence based Correlation Matrix}: \cite{ML-GCN} proposed a way to model label dependency in the form of conditional probability, i.e. $P(L_j \mid L_i)$ denotes the probability of occurrence of $L_j$ when $L_i$ is present. To construct the correlation matrix, we count the co-occurrence of label pairs present in training set to get a matrix $M \in R^{C \times C}$, where $M_{i,j}$ denotes the co-occurrence time of $L_i$ and $L_j$. We then divide $M$ by the occurrence time of $L_i$ in training set to get the conditional probability $P$. To prevent a long-tail distribution where some rare co-occurrences may be noisy, we binarize $P$ by setting a tunable threshold $t$. The correlation matrix $A$ is set to 1 if $P$ is above the $t$ and set to 0 when $P$ is below the $t$, where $A \in R^{C \times C}$.

\textbf{Ontology-based Method One}: \cite{MT-GCN} proposed the correlation matrix $A$ to denote label pairs who have same parents. $A_{i,j} = 1$ when $L_i$ and $L_j$ have same parents; $A_{i,j} = 0$ otherwise.

\textbf{Ontology-based Method Two}: \cite{MT-GCN} proposed the correlation matrix $A$ to denote label pairs who have edges in between them. In the dataset we are using, edges only occur between parents and children, so we set $A_{i,j} = 1$ when $L_i$ is a child of $L_j$ or the other way around; otherwise, $A_{i,j} = 0$.

To prevent over-smoothing, after binarizing, we re-weight $A$ to get the desired correlation matrix $A'$ by setting $A'_{i,j} = p$ when $i = j$ and $A'_{i,j} = (1-p) / \sum_{j} A_j$ when $A_{i,j} = 1$.

\subsection{Siamese Network with Graph Convolutional Network}

\begin{figure}[!ht]
    \centering
    \includegraphics[scale=0.4]{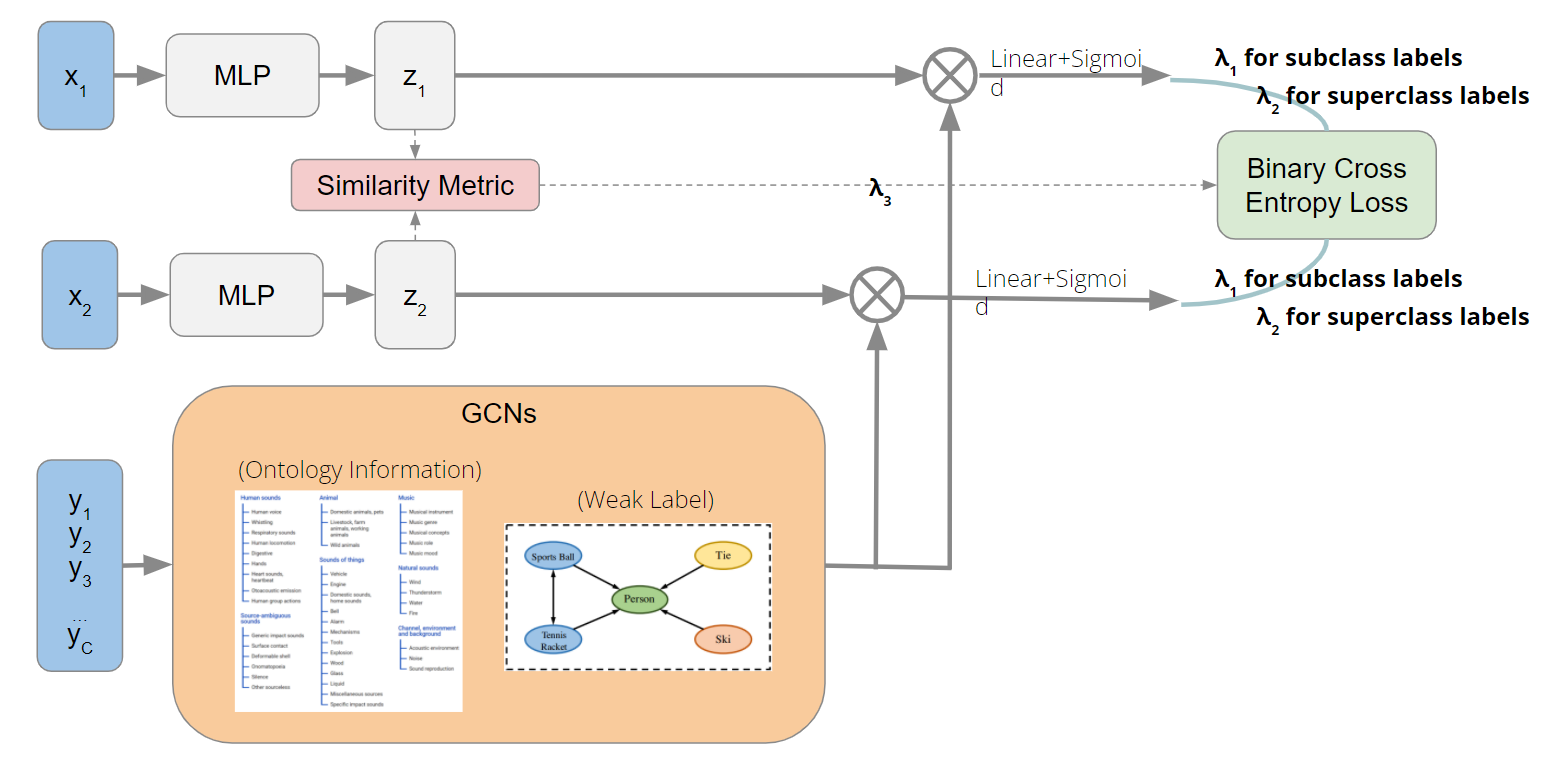}
    \caption{The framework of Siamese Network + Graph Convolutional Network }
    \label{fig:siamese_gcn}
\end{figure}

The Graph Convolution Network module described in Section \ref{gcn} can convert the label embedding from a one-hot representation to an embedding aggregating neighbor information. We use a matrix multiplication to get the similarity between a given audio clip embedding and every label's embedding. Given these similarity values, we then use a SoftMax activation and Binary Cross Entropy loss to calculate the loss between our outputs and the target labels. The overall framework is shown in Figure \ref{fig:siamese_gcn}.

Here we replace the baseline Ontological Layer with the label embeddings that the GCN generates. To combine it with Siamese Network, we still keep the $\lambda_{1}$, $\lambda_{2}$ and $\lambda_{3}$ for the loss term.  Next we will discuss the experiments in more detail and the results from the various models.


\section{Experiments}
\subsection{Evaluation Metrics}

The performance of our models is evaluated using weighted average precision and AUC from predictions.  All reported metrics are on the test set of the Audioset dataset.  As discussed previously, we extract the labels for each segment from the top two levels of the Audioset ontology. The average precision (AP) metric is first computed for each class by considering the precision (fraction of true positive labels out of all predicted positive labels) - recall (fraction of true positive labels out of actual positive labels) curve at different thresholds.  Thus it is an indication of how well the model can identify positive classes in the data.  The other metric we consider is AUC, which considers negative labels by computing the area under the TPR (true positive rate) - FPR (false positive rate) curve and is an indication of how well the model can distinguish between classes.  The AP/AUC metric for the classes on the same ontology level is combined through a weighted average based on the proportion of each class that is present in the training data to get a final weighted AP and weighted AUC score for each level.

\subsection{Performance of MLP Model with no Ontology Information}
The MLP model with no ontology information achieves a weighted AP/AUC score of 0.45099/0.8706, respectively for subclasses.  The weighted AP/AUC score for the superclasses are 0.7056/0.8556, as shown in Table \ref{tab:map_results}.  The model is trained for about 70 epochs at a learning rate of 2e-3. Please refer to the Appendix for more hyperparameter training details for each of the models.

\subsection{Performance of Siamese Network with Ontological Layer}

The Siamese model with Ontology layer achieves a subclass weighted AP/AUC = 0.3653/0.8055 and superclass weighted AP/AUC = 0.3876/0.6505. After some hyperparameter tuning of the loss function, we were able to achieve the Table \ref{tab:map_results} results with $\lambda_1 = 1.5, \lambda_2 = 1, \lambda_3=0.25$.  Through the experiments, we found that these metrics can be tuned based on the $\lambda$ values in the loss function to control the contributions from each level of the ontology.  A more detailed table of the hyperparameter tuning is presented in the appendix, and the results in the summary table are chosen as a good balance between the two ontology levels.  

This experiment shows there should be more research work to apply a Siamese network to a multi-label scenario, such as how to generate pairs of the same sub/superclass as well as their similarity metric.  We further investigated this problem by considering two definitions: one in which a pair of data is of the "same" subclass if they have exactly the same labels and another in which "same" means that two pairs just have some intersection in their labels. Our experiments suggest that the latter definition is better for the multi-label scenario.  However, the performance of this baseline Siamese model which uses ontology information suggests that this external knowledge is still difficult to embed into a model for weakly labeled data. Furthermore, we believe that this definition of "same" or different" sub/superclass is still ambiguous and introduces even more noise into the model as it attempts to cluster pairs that are not really the "same" subclass. This can explain why we see such poor performance metrics compared to the MLP with no ontology information.

\begin{table}[!ht]
    \centering
    \begin{tabular}{c|c|c|c|c} 
                            & \multicolumn{2}{c|}{Weighted AP} & \multicolumn{2}{c}{Weighted AUC} \\ \cline{2-5}
         Model              & Subclass Level & Superclass Level & Subclass Level & Superclass Level \\ \hline
         MLP                & 0.4509 & 0.7056 & 0.8706 & 0.8556 \\
         Siamese + Ontology & 0.3653 & 0.3876 & 0.8055 & 0.6505 \\
         Siamese + GCN      & 0.4285 & 0.6790 & 0.8460 & 0.8280 \\
         MLP + GCN          & \textcolor{red}{0.4590} & \textcolor{red}{0.7117} & \textcolor{red}{0.8751} & \textcolor{red}{0.8602} \\
   \end{tabular} 
   \vspace{10pt}
    \caption{mAP and AUC Results of different models}
    \label{tab:map_results}
\end{table}


\begin{figure}[ht]
\centering
\includegraphics[width=\textwidth]{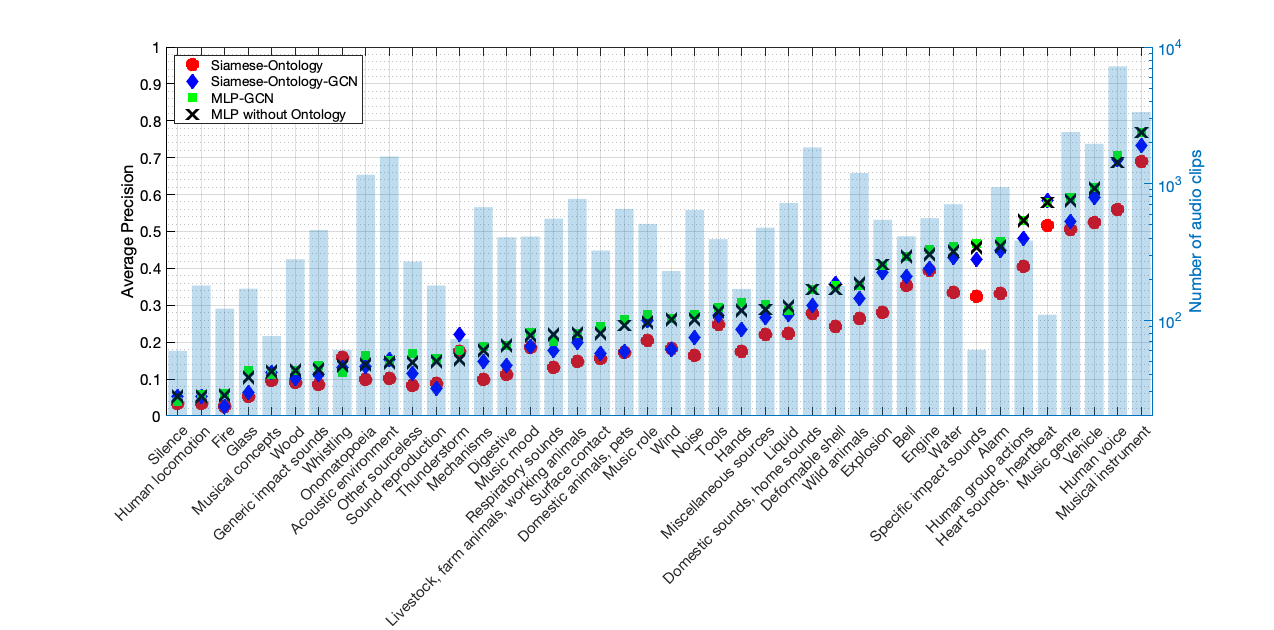}
\caption{mAP across different low level labels}
\label{fig:map_subclass}
\end{figure}

\begin{figure}[ht]
\centering
\includegraphics[width=\textwidth]{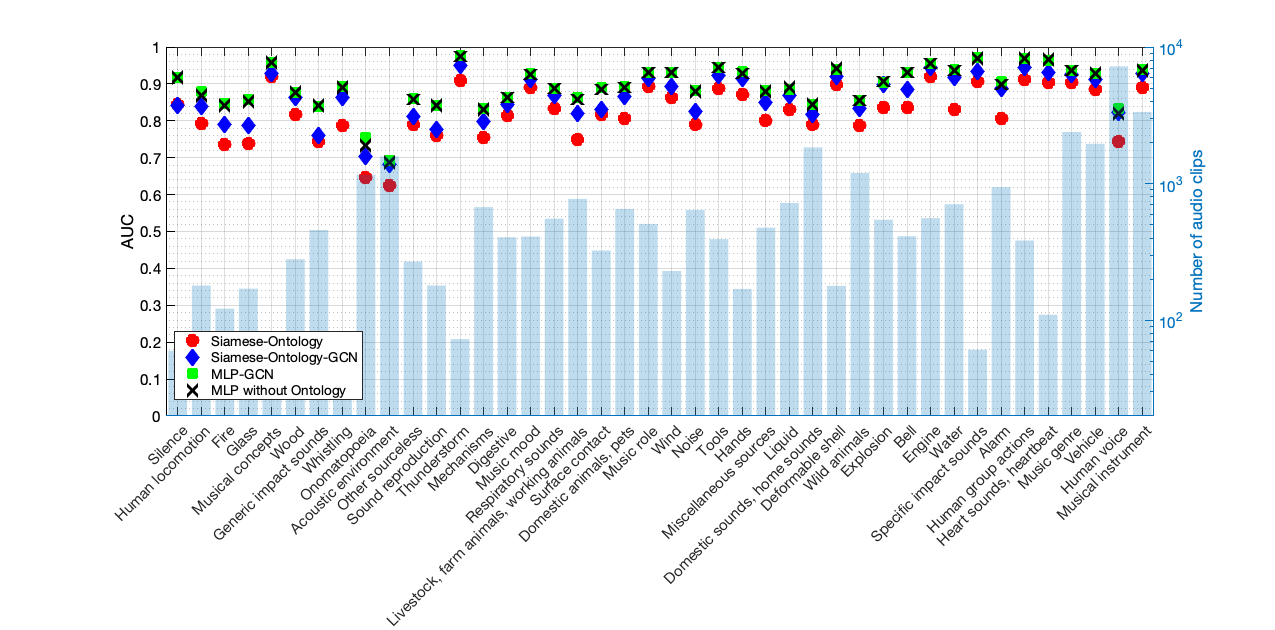}
\caption{AUC across different low level labels}
\label{fig:auc_subclass}
\end{figure}


\begin{figure}[ht]
\begin{minipage}[b]{0.49\textwidth}
    \centering
    \includegraphics[scale=0.25]{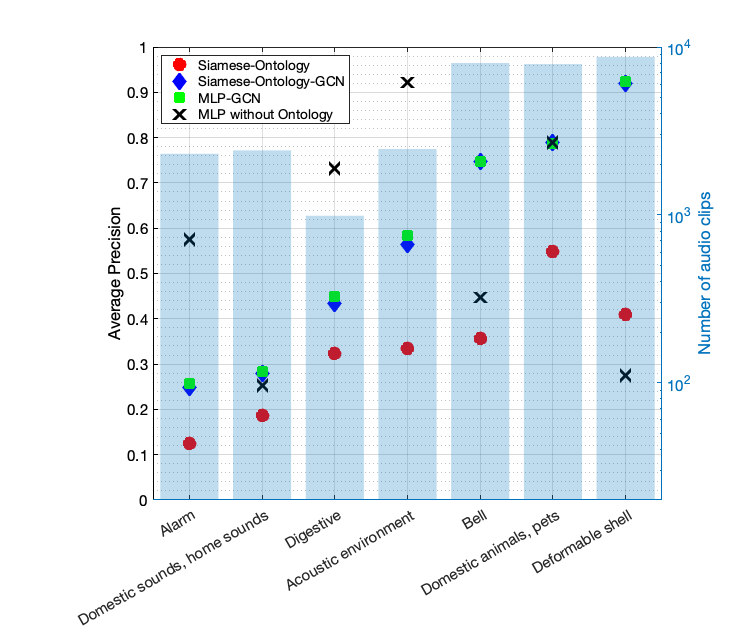}
    \caption{AP across different superclass labels}
    \label{fig:ap_superclass}
\end{minipage}
\begin{minipage}[b]{0.49\textwidth}
    \centering
    \includegraphics[scale=0.25]{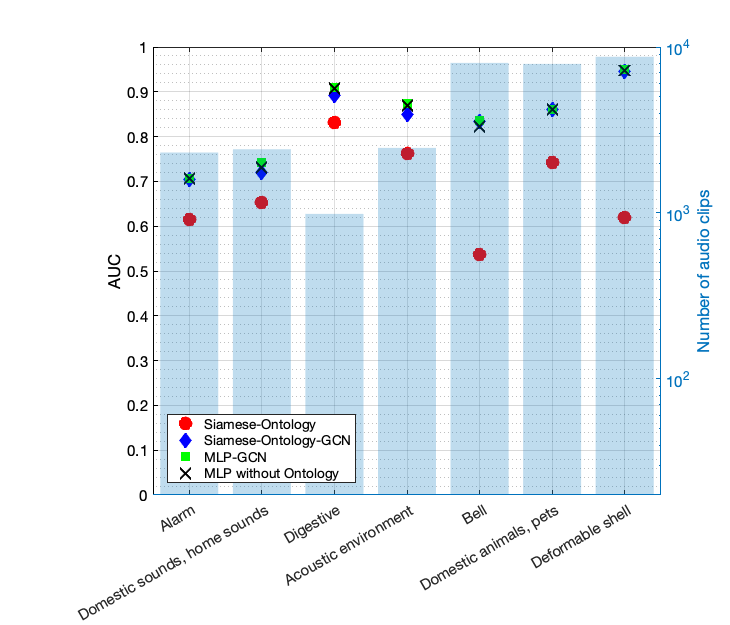}
    \caption{AUC across different superclass labels}
    \label{fig:auc_superclass}
\end{minipage}
\end{figure}

\subsection{Performance of Siamese-GCN Model}

The extension of the Siamese model replaces the Ontology layer with a Graph Convolution network to embed ontology information. We trained the model by Adam optimizer for about 30 epochs. We use 2 Linear Layer as the node embedding of GCN and apply 2 Layer GCN. The performance is sensitive to the hyper-parameters $\lambda_{1}$, $\lambda_{2}$ and $\lambda_{3}$ of the loss function because it would amplify or reduce learning rate for different loss terms. The Table \ref{tab:siamese_gcn_results} shows how different hyper-parameters could affect the performance. One interesting observation is that the greater $\lambda$ is not necessary equal to the greater learning rate.

In this experiment, we also can see that the Siamese network with GCN could tolerate the noises introduced from weakly labelled data better than the baseline Siamese with an un-trainable Ontological Layer.  This suggests that the GCN can better capture the ontology hierarchy in the model and that it can even overcome the noise introduced by attempting to find pairs for the Siamese net.  Overall, the evaluation metrics for this model are close to the baseline MLP with no ontology information. 

\subsection{Performance of MLP-GCN Model}

To study the utility of the Siamese net we also implement an MLP-GCN using the same GCN structure as the one in Siamese GCN framework.  We also take a deeper look into which correlation matrices can provide the best improvement in performance. Among the 3 correlation matrices, we found labels co-occurrence based correlation matrix works best to fit the weak label scenario and model the ontology information. Thus, we furthermore did more experiments on the two trainable parameters: $t, p$ (see section 3.5.1).

From our experiments (see Appendix A.4), we found $p=0.2, t=0.08$ achieved the best results. We could see that although this model performs better than the previous models, the improvement from MLP without Ontological Layer is still limited. This suggests that even with the ontology information embedded into the GCN, it is still hard to get significant improvement in classification results. It also further demonstrates that the architecture of the Siamese net is not a great fit for this multi-label scenario.  By using a simple MLP network to learn embeddings, it can already achieve relatively good performance with the GCN.  Overall, we found that there were certain classes for which it was always difficult to achieve high AP or AUC scores as demonstrated in Figures \ref{fig:map_subclass}, \ref{fig:auc_subclass}, \ref{fig:ap_superclass}, \ref{fig:auc_superclass}. For these classes, such as glass, fire, or silence, we analyzed which data points contain those classes and found that they are often multi-labeled with more commonly found classes in the training set, such as human voice, or domestic sounds.  This could make it difficult for the model to learn different representations for those classes.

\section{Conclusions}

To conclude, we observed that the Siamese model with ontology layer has the worst performance, while simple MLP is getting relatively better results, and the combination of MLP with GCN works even better. It is possible for the Siamese architecture to have a negative impact on prediction, introducing even more confusion through the construction of input pairs for data that is weakly multi-labeled.  Although the GCN provides some slight improvement, it seems that the model still has difficulty in differentiating ambiguous subclasses or using the hidden knowledge in weakly labeled data. The way to incorporate ontology information might be different with respect to contexts (dataset, ontological architecture, network structure, etc.), and it would take a lot more research effort to determine the best ways to use it. For a dataset like AudioSet, where each data point has multiple, or even wrong labels, having a simple ontology layer on the output end of a Siamese network might not be a good choice. So the next step might be to narrow down the scope of data-set and try different approaches to embed ontological information and approaches that could also make use of deeper levels of ontology information.

\medskip

\small
\bibliographystyle{IEEEtranN}
\bibliography{midterm}

\newpage
\appendix
\section{Hyperparameters}
\subsection{MLP Model with no Ontology Information}
The simple MLP model is trained using the Adam optimizer with a learning rate=$2e-3$ and weight decay=$1e-4$, trained for around 73 epochs. 

\subsection{Siamese with Ontology Layer}
\begin{table}[!ht]
    \centering
    \begin{tabular}{c|c|c|c|c|c|c} 
        $\lambda_1$ & $\lambda_2$ & $\lambda_3$ & Low Level mAP & High Level mAP & Low Level AUC & High Level AUC \\ \hline
        1.5  & 1 & 0.25 & 0.3379 &	\textcolor{red}{0.4470} &	0.7904 &	\textcolor{red}{0.7061} \\
        1.75 & 1 & 0.25 & 0.3480 &	0.4357 &	0.7941 &	0.6959 \\
        2 & 1 & 0.25 & {0.3511} & 0.4175 & 0.7969	& 0.6799 \\
        2 & 0.75 & 0.5 & \textcolor{red}{0.3653} &	0.3876 &	\textcolor{red}{0.8055} &	0.6505 \\
        2 & 1 & 0.5 & {0.3586} & 0.4196 &	0.8004 & 0.6845 \\
    \end{tabular}
    \vspace{10pt}
    \caption{mAP and AUC Results of MLP-GCN with different $p$ parameters when $t=0.168$}
    \label{tab:Siamese_results}
\end{table}

\subsection{Siamese-GCN}
\begin{table}[!ht]
    \centering
    \begin{tabular}{c|c|c|c|c|c} 
        $\lambda_{1}$ & $\lambda_{2}$ & $\lambda_{3}$ & lr & Low Level mAP & High Level mAP \\ \hline
        20 & 2 & 0.5 & $1e-3$ & 0.34 & 0.625 \\
        100 & 100 & 0.5 & $1e-3$ & \textcolor{red}{0.4285} & \textcolor{red}{0.679} \\
        1 & 1 & 0.005 & $1e-1$ & 0.306 & 0.615 \\
    \end{tabular}
    \begin{tabular}{c|c|c|c|c|c} 
        $\lambda_{1}$ & $\lambda_{2}$ & $\lambda_{3}$ & lr & Low Level AUC & High Level AUC \\ \hline
        20 & 2 & 0.5 & $1e-3$ & 0.796 & 0.788 \\
        100 & 100 & 0.5 & $1e-3$ & \textcolor{red}{0.846} & \textcolor{red}{0.828} \\
        1 & 1 & 0.005 & $1e-1$ & 0.797 & 0.772 \\
    \end{tabular}
    \caption{mAP and AUC results of Siamese GCN with different $\lambda$ and learning rate (lr)}
    \label{tab:siamese_gcn_results}
\end{table}

\subsection{MLP-GCN}
\begin{table}[!ht]
    \centering
    \begin{tabular}{c|c|c|c|c|c} 
        p & Low Level mAP & High Level mAP & Low Level AUC & High Level AUC \\ \hline
        $0.1$ & 0.4477 & 0.7091 & 0.8571 & \textcolor{red}{0.8690} \\
        $0.2$ & \textcolor{red}{0.4469} & \textcolor{red}{0.7098} & \textcolor{red}{0.8693} & 0.8575 \\
        $0.3$ & 0.4467 & 0.7083 & 0.8568 & \textcolor{red}{0.8690} \\
    \end{tabular}
   
    \caption{mAP and AUC Results of MLP-GCN with different $p$ parameters when $t=0.168$}
    \label{tab:MLP-GCN_results}
\end{table}

\begin{table}[!ht]
    \centering
    \begin{tabular}{c|c|c|c|c|c} 
        t & Low Level mAP & High Level mAP & Low Level AUC & High Level AUC \\ \hline
        $0.010$ & 0.4455 & 0.7072 & 0.8687 & 0.8566 \\
        $0.025$ & 0.4479 & 0.7077 & 0.8694 & 0.8567 \\
        $0.080$ & \textcolor{red}{0.4501} & \textcolor{red}{0.7141} & \textcolor{red}{0.8710} & \textcolor{red}{0.8613} \\
        $0.100$ & 0.4471 & 0.7128 & 0.8696 & 0.8597 \\
        $0.150$ & 0.4462 & 0.7093 & 0.8691 & 0.8577 \\
    \end{tabular}
   
    \caption{mAP and AUC Results of MLP-GCN with different $t$ parameters when $p=0.2$}
    \label{tab:MLP-GCN_results}
\end{table}

\end{document}